\documentclass[manuscript,screen]{acmart}
\AtBeginDocument{%
  \providecommand\BibTeX{{%
    \normalfont B\kern-0.5em{\scshape i\kern-0.25em b}\kern-0.8em\TeX}}}

\setcopyright{acmlicensed}
\copyrightyear{2024}
\acmYear{2024}
\acmDOI{XXXXXXX.XXXXXXX}
\acmConference[With or Without Permission at CHI '24]{Site-specific Augmented Reality for Social Justice}{May 12,
  2024}{Honolulu, HI}
\acmBooktitle{With or Without Permission at CHI '24: Site-Specific Augmented Reality for Social Justice,
 May 12, 2024, Honolulu, HI} 
\acmISBN{978-1-4503-XXXX-X/18/06}

\begin{document}

% \DeclareRobustCommand{\okina}{%
%   \raisebox{\dimexpr\fontcharht\font`A-\height}{%
%     \scalebox{0.8}{`}%
%   }%
% }
% \newunicodechar{ʻ}{\okina}

\title{Revealing Aspects of Hawai'i Tourism Using Situated Augmented Reality }

\author{Karen Abe}
\affiliation{%
  \institution{Independent, Honolulu, HI}
  \city{Honolulu} 
  \state{Hawai'i} 
  \country{USA}
}
\email{prod.karenabe@gmail.com}

\author{Jules Park}
\affiliation{%
  \institution{York University, Toronto, ON}
  \city{Toronto} 
  \state{Ontario} 
  \country{Canada}
}
\email{hjoo@my.yorku.ca}

\author{Samir Ghosh}
\affiliation{%
  \institution{University of California, Santa Cruz, CA}
  \city{Santa Cruz} 
  \state{California} 
  \country{USA}
}
\email{samir.ghosh@ucsc.edu}


\begin{abstract}
    In this position paper, we present a process artifact that aims to bring awareness to historical context, contemporary issues, and identity harm inflicted by tourism in Hawai‘i. First, we introduce the historical background and how the work is informed by the positionality of the authors. We discuss how related augmented reality work can inform strategy for building augmented reality experiences that address cultural issues. Then, we present a mockup of the artifact, aimed to bring awareness to 20th-century colonialism, recent Kānaka Maoli art exclusion, and cultural prostitution.  We describe how we will share the app at the workshop and list topics for discussion.
\end{abstract}



% \begin{figure}[h!] 
%   \includegraphics[width=\textwidth]{process-artifacts-images/Teaser.png}
%   \caption{Teaser image depicting a site-specific historical artifact situated in mobile augmented reality (MAR)}
%   % \Description{In this mockup, a user }
%   \label{fig:mockup1}
% \end{figure}

\begin{teaserfigure}
  \includegraphics[width=\textwidth]{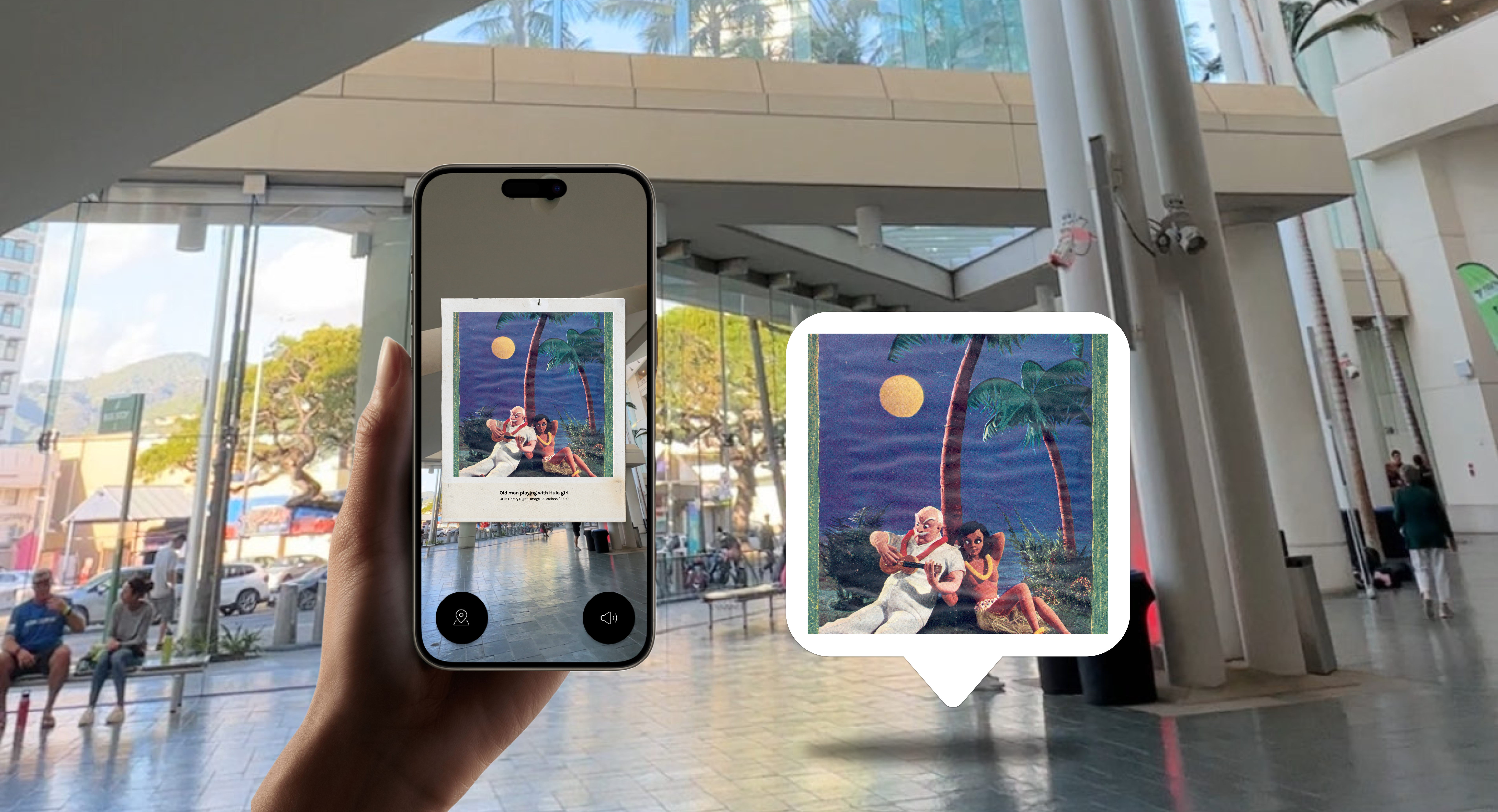}
  \caption{Mockup depicting a site-specific historical artifact situated in mobile augmented reality}
 \Description{Mockup depicting a site-specific historical artifact situated in mobile augmented reality}
  \label{fig:teaser}
\end{teaserfigure}

\received{8 March 2024}

\maketitle

\section{Introduction}

Hawai‘i is often perceived as a “paradise” by tourists, offering beautiful beaches, water activities, and “exotic” culture. However, this image of “paradise” masks the struggles of Hawai'i locals and the Kānaka Maoli community who face high living costs and social injustices. (Note: Kānaka Maoli, the indigenous people of Hawai'i. Often referred to as "Kānaka" for short.) Upon the overthrow of the Hawaiian Kingdom in 1893, white politicians and businessmen spearheaded the plan for building Hawai‘i as a tourist destination. In a working paper for The Economic Research Organization at the University of Hawai‘i, James Mak describes the history of colonial tourism, referring to a 1902 publication by the Merchants’ Association of Honolulu where the committee noted their hopes of expanding tourism to foster a “permanent population of the most desirable character” and shape Hawai‘i into a “white republic" \cite{mak_creating_2015}. To this day, the drivers and beneficiaries of the tourism industry are controlled by non-Kānaka politicians and business owners, thus the historical context of tourism and its colonialist intentions are often hidden from tourists and visitors. \cite{abe_home_2023} These perceptions often influence contemporary social justice issues such as the recent exclusion and reconciliation of Kānaka Maoli art in the Hawaii Convention Center \cite{lincoln_controversial_2013}. Furthermore, this colonial tourism diminishes the cultural values and identities of the Kānaka Maoli people. Haunani-Kay Trask refers to the practice of the commercialization of altered native values and practices as 'cultural prostitution' which leads to cultural exploitation and erasure \cite{trask_lovely_1999}.

Situated augmented reality offers an opportunity to bring digital assets that depict and represent these issues: (1) Historical Context, (2) Contemporary Issues, and (3) Cultural Prostitution, in front of tourists who are actively inhabiting Hawaiian space.

We reflect on our positionality and its potential impact on how we create artifacts that relate to culture and identity. The first author is a Hawai‘i-grown media artist with experience in serving the Hawai‘i State legislature and grass-roots advocacy groups. Their research-based projects explore issues around settler colonialism, militarism, and tourism as a way to challenge the dominant historical narratives of Hawai‘i. The first author, although born and raised in Hawai‘i is not of Kānaka Maoli or Native Hawaiian descent. Thus, is navigating their work as a settler colonist on the lands of the Kānaka Maoli people. The second author and third authors are contributors who identify as Asian-American and consider how their non-native identity should relate to situated augmented reality work.

\section{Related Work}

Hawaiian Airlines recent implementation of the “Travel Pono” in-flight announcement video \cite{hawaiian_airlines________________travel_2021} is an example of (non-XR) media strategies employed by businesses to educate the audience on the impact of tourism and present best practices for visitors. The video was crafted in a way where Hawaiian airlines staff — flight attendants, pilots to engineering staff — individually introduce aspects of ways to travel “Pono” or responsibly. The video successfully presents topics of sustainability and cultural preservation through the voices of the “local community.” 

For the depiction of archival materials, Booksnake is an app developed by digital humanities scholars that brings archival material from institutions such as the Huntington Library and Library of Congress into augmented reality environments as manipulable assets.\cite{fraga_booksnake_2023}. In this way, Booksnake demonstrates how augmented reality empowers and permits users greater access to materials from archival institutions. In a situated augmented reality app, these historical assets could be collocated with the present day, to encourage tourists to engage with the historical layer of their inhabited space.

For aspects of reclaiming historical narrative, Kinfolk shows myriad approaches in using AR to inform the user about marginalized Black and Brown histories \cite{brewster_kinfolk_2017}. By drawing upon these techniques, social justice applications for situated augmented reality could describe the marginalization of Hawaiian experiences due to colonial tourism.

\section{Process Artifact}

% \begin{figure}
%   \includegraphics[width=\textwidth]{process-artifacts-images/AR-Mockup.png}
%   \caption{In this mockup, a user views a historical artifact and context using situated augmented reality}
%   % \Description{In this mockup, a user }
%   \label{fig:mockup1}
% \end{figure}

% \begin{figure}[h!]
%   \includegraphics[width=\textwidth]{process-artifacts-images/Hula-Mock-Up.png}
%   \caption{In this mockup, a user views a historical artifact and context using situated augmented reality}
%   % \Description{In this mockup, a user }
%   \label{fig:mockup1}
% \end{figure}

\begin{figure}[h!]
  \includegraphics[width=\textwidth]{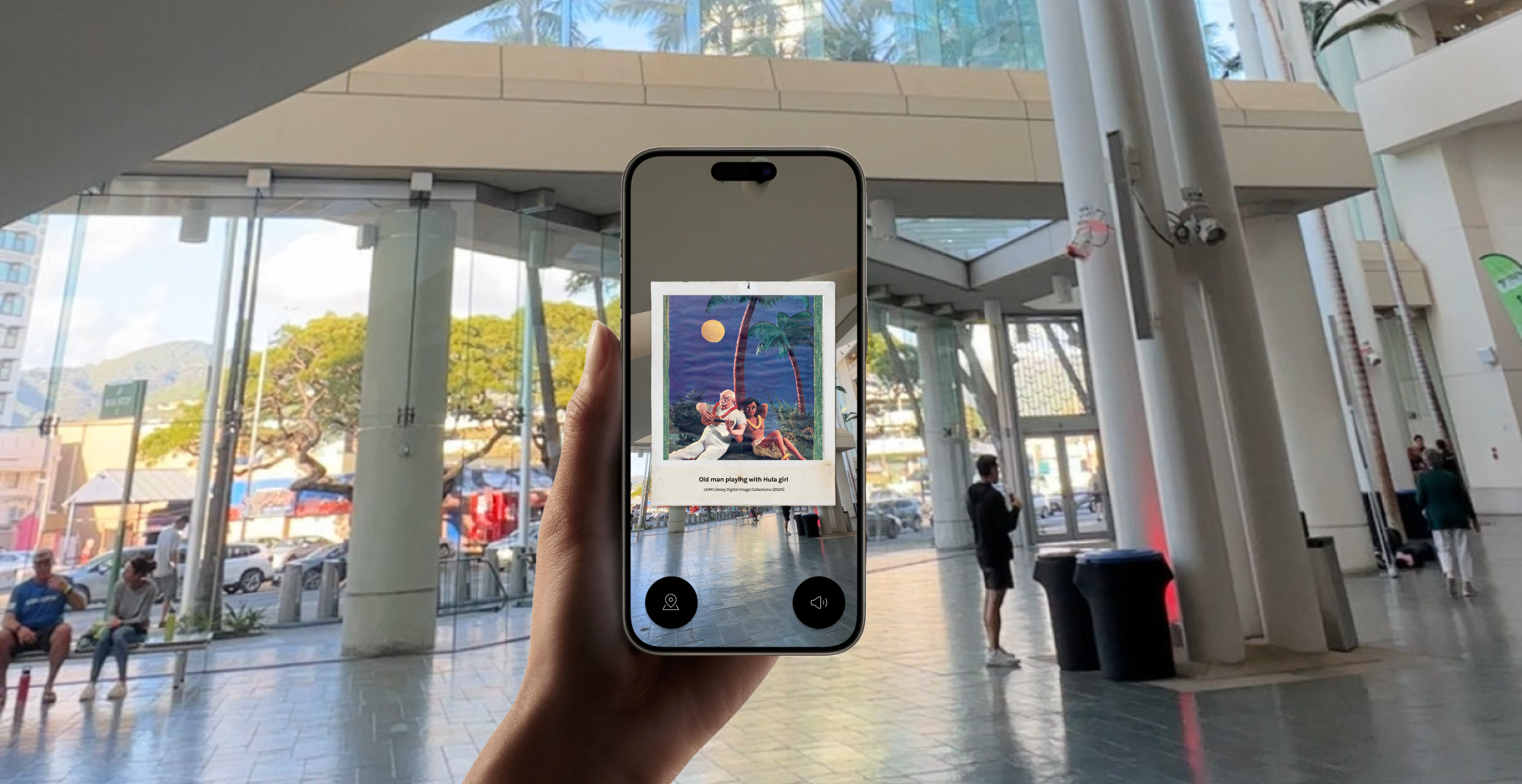}
  \caption{In this mockup, a user views a historical artifact and context using situated augmented reality}
  \Description{In this mockup, a user is scanning the environment with their mobile phone and sees a augmented reality panel representing a historical artifact}
  \label{fig:mockup1}
\end{figure}

In our process artifact, we intend to display assets that address the following aspects of the effect of colonial tourism:
\begin{itemize}
\item \textbf{Historical Context} - Provide information about the history, and intention of settlers in Hawai‘i through the display of historical artifacts that depict scenes from the late 19th and early 20th century.
\item \textbf{Contemporary Issues} - Bring awareness to recent issues regarding tourism. For example, we can show news articles or secondary sources on the recent exclusion of art in the convention center \cite{lincoln_controversial_2013}.
\item \textbf{Identity Harm} - Display examples of commercialized or harmful caricatured native values or practices. In Fig.\ref{fig:mockup1}, we place a 1943 advertisement of a sexualized "Hawaiian" hula girl singing alongside a white man, as an example of cultural prostitution of the native practice of hula. \cite{warner_page_1943}
\end{itemize}

\begin{figure}
  \includegraphics[width=\textwidth]{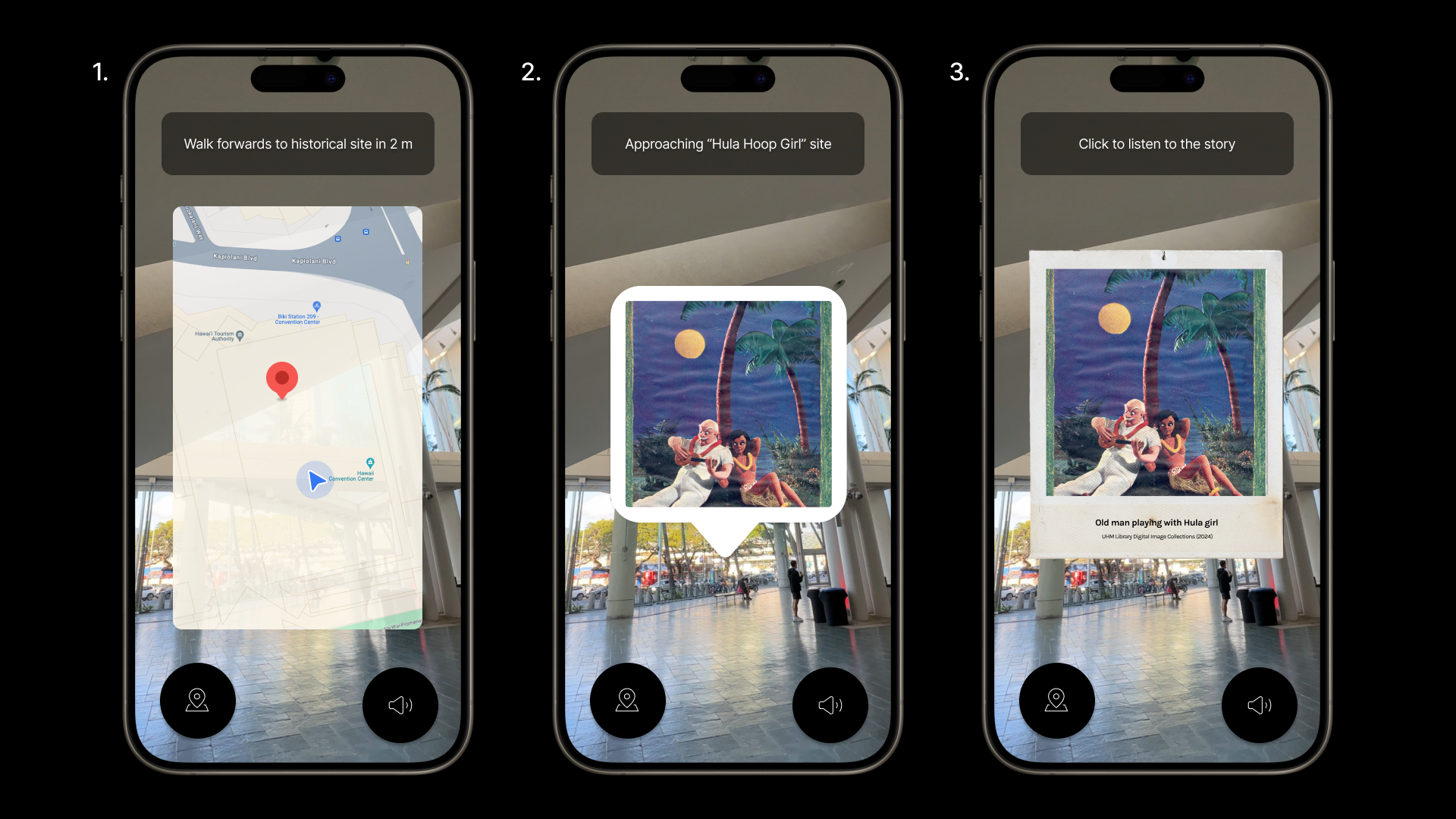}
  \caption{A user interface conceptually depicting a wayfinding system for navigation in situated augmented reality in a real world setting}
  \Description{In this image, a wayfinding design system is represented showcasing how the augmented reality experience will take place across site-specific locations}
  \label{fig:mockup2}
\end{figure}

The prototype will showcase an augmented reality (AR) storytelling experience that leverages cutting-edge technology enhanced with geolocation in order to design site-specific AR interactions. Users are invited to engage with a variety of public visual position system (VPS) locations featuring digital AR objects placed across the convention center. By physically visiting these sites and utilizing their smartphones to scan the designated markers, these markers will contextually transform these sites with augmented historical artifacts embedded within the fabric of the physical environment. The core technological backbone powering this experience will be designed using the 8th Wall Lightship SDK combined with the VPS World Explorer system. These tools will allow a robust AR experience, accessible to the public through mobile devices. By directly interacting with these site-specific AR artifacts, users will experience a variety of stories depicting critical topics in both the historical and contemporary Hawaiian cultures communicating information about settler history. The stories and critical events represented in this experience will be collected and curated from real community members from the Hawai'ian community. AR artifacts leveraged with VPS geolocation technology will build a seamless, site-specific experience connecting the user with both the digital and real-world environments. Users will be encouraged to explore and interact with these digital artifacts to learn more information, navigate to other nearby AR artifacts available on a map interface, and immerse themselves through AR portals into a 3D environment depicting the story. 

The artifacts presented through AR will be collected through engagement of the Kānaka Maoli and local Hawai'i community. From conversations on key narratives, to showcasing artwork produced by contemporary Kānaka Maoli artists, this artifact will push for greater representation of Kānaka Maoli and local Hawai'i voices. We intend for the AR medium to serve as a tool to amplify local and native concerns to the greater audience.

\section{Topics for Discussion}

\begin{itemize}
\item How do we best present and contextualize archival materials for social justice?
\item How do ensure that situated augmented reality empowers locals?
\item What are best practice for facilitating local institutions' augmented reality capabilities?
\end{itemize}

\bibliographystyle{ACM-Reference-Format}
\bibliography{CHI-AR-24}

% %%
% %% If your work has an appendix, this is the place to put it.
% \appendix

\end{document}